\documentclass[12pt,onecolumn,manuscript]{article}
\usepackage[dvips]{graphicx}
\usepackage{epsfig, color}
\usepackage{pstricks}
\usepackage{amssymb,amsfonts,amsmath}

\providecommand{\R}{\mathbb{R}}
\providecommand{\X}{\mathcal{X}}

\begin{document}

\title{Topological Methods for Exploring Low-density States on Biomolecular Folding Pathways}



\author{
Yuan Yao\footnote{Department of Mathematics, Stanford University, Stanford, CA 94305. Email: {\tt yuany@stanford.edu}. }~~~
Jian Sun\footnote{Department of Computer Science, Stanford University, Stanford, CA 94305. Email: {\tt sunjian@stanford.edu}.}~~~
Xuhui Huang\footnote{Department of Bioengineering, Stanford University, Stanford, CA 94305. Email: {\tt huangx@stanford.edu}.}~~~
Gregory R. Bowman\footnote{Biophysics Program, Stanford University, Stanford, CA 94305. Email: {\tt gbowman@stanford.edu}.}~~~
Gurjeet Singh\footnote{Department of Mathematics, Stanford University, Stanford, CA 94305. Email: {\tt gurjeet@stanford.edu}.}\and
Michael Lesnick\footnote{Insititute of Computational and Mathematical Engineering. Email: {\tt mlesnick@stanford.edu}.}~~~
Leonidas J. Guibas\footnote{Department of Computer Science, Stanford University, Stanford, CA 94305. Email: {\tt guibas@cs.stanford.edu}.}~~~
Vijay S. Pande\footnote{Department of Chemistry, Stanford University, Stanford, CA 94305. Email: {\tt pande@stanford.edu}.}~~~
Gunnar Carlsson\footnote{Department of Mathematics, Stanford University, Stanford, CA 94305. Email: {\tt gunnar@math.stanford.edu}.}}

\maketitle

\newpage 

\begin{abstract} 
\noindent {\bf Characterization of transient intermediate or transition states is crucial for the description of biomolecular folding pathways, which is however difficult in both experiments and computer simulations.  Such transient states are typically of low population in simulation samples. Even for simple systems such as RNA hairpins, recently there are mounting debates over the existence of multiple intermediate states. In this paper, we develop a computational approach to explore the relatively low populated transition or intermediate states in biomolecular folding pathways, based on a topological data analysis tool, Mapper, with simulation data from large-scale distributed computing. The method is inspired by the classical Morse theory in mathematics which characterizes the topology of high dimensional shapes via some functional level sets. In this paper we exploit a conditional density filter which enables us to focus on the structures on pathways, followed by clustering analysis on its level sets, which helps separate low populated intermediates from high populated uninteresting structures. A successful application of this method is given on a motivating example, a RNA hairpin with GCAA tetraloop, where we are able to provide structural evidence from computer simulations on the multiple intermediate states and exhibit different pictures about unfolding and refolding pathways. The method is effective in dealing with high degree of heterogeneity in distribution, capturing structural features in multiple pathways, and being less sensitive to the distance metric than nonlinear dimensionality reduction or geometric embedding methods. The methodology described in this paper admits various implementations or extensions to incorporate more information and adapt to different settings, which thus provides a systematic tool to explore the low density intermediate states in complex biomolecular folding systems.}
\end{abstract}



\section*{Author Summary}
In biomolecular folding problems, identification of intermediate or transition states connecting folded or unfolded states is crucial for understanding the folding pathways, but is difficult in both experiments  due to the transient nature and computer simulations due to the low sampling population. We are motivated by recent debates over whether RNA hairpins fold in a two-state or multi-state manner with intermediates on pathways. To address this issue, we develop a computational approach to explore the relatively low populated transition or intermediate states in biomolecular folding pathways, based on a topological data analysis tool, Mapper, with simulation data from large-scale distributed computing. The method in this paper exploits a conditional density filter which effectively focuses on pathways, followed by clustering analysis on filter level sets which separates low populated intermediate states from others of high population. Such a scheme, when applied to a RNA hairpin system, successfully provides important structural evidence on the multiple intermediate states on folding pathways which may guide further experimental investigations. The methodology adds to the communities of simulation and informatics an exploratory tool to data mine large data sets and propose experimentally testable properties, such as folding pathways for biological macromolecules. 


\section{Introduction}
The folding of biomolecules is a classic biophysical problem. 
Proteins and nucleic acids are synthesized as linear polymer chains. 
They must then spontaneously and rapidly fold into their three-dimensional native states. 
The folding process is determined by the underlying free energy landscape.
These landscapes are rugged, having many local minima corresponding to intermediates and misfolded states. Characterizing these states is critical for a full understanding of biomolecular folding.
Experimental studies may point to the existence of such states but are usually unable to provide high resolution structural information due to the transience and/or heterogeneity of such states.
Computer simulations have proved useful for sampling this complex high-dimensional space while yielding structures at full-atom resolution.
However, these simulations tend to generate millions of configurations.
The volume and high-dimensional nature of the output make it extremely difficult to discern the structure of the data.

One common approach to dealing with computer simulation results is to apply K-means clustering to the entire data set.
However, K-means clustering suffers from a number of important limitations.
First, it is limited by the need to specify the number of states from the beginning. Second, it tends to create spherical states. The relevant states of the free energy landscape, on the other hand, may be non-convex. In this case, K-means clustering will tend to lump unrelated configurations together or split related configurations into separate states. This limitation may be overcome by splitting the configurations into many small states and grouping them together using various metrics that allow non-convex states, such as in \cite{Chodera07}. There is another widely used clustering method, single-linkage, which may overcome these issues in K-means.
Unfortunately, identifying sparsely populated intermediate states is still difficult. 
Simulation data tends to be heavily dominated by the most stable states, such as the folded and unfolded states, and single-linkage clustering of the entire data set tends to pick up densest states only and hardly distinguish the intermediates from noise.

Recently, geometric embedding techniques, such as nonlinear dimensionality reduction [2-7], have been explored as a means to overcome the dimensionality hurdle in complex biomolecular systems.
For example, ISOMAP \cite{TDL00} has been applied to protein folding \cite{KavPNAS06} and Laplacian eigenmap \cite{BN03} has been applied to the dynamics of biological networks \cite{GG07}. 
This class of techniques maps the data in high dimensional spaces to a low dimensional space by preserving some local/global metric relationship among neighboring data points. In this way, one can easily visualize data and possibly gain important insights. For instance, the new embedding coordinates may be biologically relevant reaction coordinates \cite{KavPNAS06}. 
However, the performance of these geometric embedding techniques will suffer from the high degree of heterogeneity in distribution and be sensitive to the choice of the distance metric. 

One efficient strategy to address these issues is to stratify the data into density level sets and study its topological features such as clustering which are less sensitive to the metric than geometric methods.
High density levels will contain the dominant states, such as the folded and unfolded states, while less populated states, such as intermediates, will occupy the low density levels. Clustering on level sets of similar density, will be less affected by the distributional heterogeneity and thus effectively disclose structural information about intermediates. 
This idea of stratification is reminiscent of Morse theory, which provides a general machinery for studying the topology of high dimensional manifolds by looking at level sets of some nicely-behaved function \cite{Milnor63}. Inspired by Morse theory, Singh {\it et al.} recently introduced Mapper \cite{SMC07}, a topological data analysis tool for high dimensional data sets. 

Mapper is a way to visualize and cluster high dimensional data. In its simple form, a filter function is used to decompose the data into overlapping level sets and clustering is then carried out in each of them. 
A graph is then generated by connecting clusters in neighboring level sets with an edge if they have non-empy overlapping. If an energy function is taken as the filter, such a graph in the limiting case will summarize the same kind of topological information as a disconnectivity graph of the energy landscape \cite{BecKar97}. But the flexibility of the filter design in Mapper makes it applicable in non-equilibrium data as well. In its extended form, Mapper can return a simplicial complex with high dimensional topological information about the data. The method is computationally efficient and amenable to parallelization. 


In this work we demonstrate the applicability of Mapper in its simple form to the biomolecular folding problem.
We begin with a discussion of Mapper itself from a perspective of Morse Theory and then present the details of a filtering function that is well-suited for biomolecular folding problems, the conditional density filter. 
This filter weights configurations close to a state of interest more heavily, thus facilitating the identification of any intermediate state leading up to it.
We then describe the use of single-linkage clustering within level sets to allow the identification of an unspecified number of non-convex states.
Finally, we discuss the application of Mapper to the folding of a small RNA hairpin, which gives some structural evidence from computer simulations in support of the multi-state hypothesis \cite{Ma07}.
The biological implications of the Mapper results are discussed by \cite{Greg08} elsewhere.
We also briefly discuss the advantages of Mapper over nonlinear dimensionality reduction techniques. 
In the future we hope to explore the combination of those geometric embedding techniques with Mapper in order to take advantage of the strengths of both approaches.

\section{Methods}

\subsection{Mapper: A Tool for Topological Data Analysis}

One way to reduce the computational complexity in the study of massive data sets is to decompose the data by classifying the data into groups and doing analysis on each of the group individually instead of performing analysis on the whole. This strategy is amenable for parallel computation, which is particular important for studies of biomolecular folding, where a great amount of configurations are normally generated.

Here we pursue this idea in the particular case where the decomposition is induced by the choice of some filter function on the data set, $h: \X\to \Omega$. In this paper, we will only consider filters that take values in the real line, though the Mapper methodology is equally applicable for filter functions taking values in higher dimensional space, or even spheres, tori, or any other topological space. With this choice, we introduce Mapper from a perspective of Morse Theory, which differs from the original paper \cite{SMC07} but discloses a deeper inspiration. 

\emph{Morse Theory} \cite{Milnor63} tells us that when $h:\X\to \R$ is some nicely-behaved function, topological information of $\X$ can be inferred from the level sets $h^{-1}(\omega)$. Such nice functions are called Morse functions in mathematics, which are those smooth functions with only nondegenerate critical points -- in other words, the Hessian at each critical point where the gradient vanishes has full rank. Morse functions are generic in the sense that they are dense in the space of smooth functions, as well of continuous functions. Hence every continuous function can be approximated arbitrarily well by Morse functions. Morse theory is an extremely powerful tool to analyze the topology of high dimensional manifolds, which lies in the heart of proving the celebrated Poincare Conjecture of dimension no less than five \cite{Smale61}. 

The simplest example in this spirit may be \emph{Reeb graph} \cite
 {Ree46}, by contracting to points the connected components within level sets $h^{-1}(\omega)$, illustrated as Figure 1 (a). This simple scheme turns out to be useful in various fields under different names, {\it e.g.} contour trees in computational geometry \cite{KrevOos97} and cluster trees 
in statistics \cite{Hart81,Stuetzle03,ZhouWong08}.  

{\it Mapper} \cite{SMC07} extends this construction to incorporate the discrete setting where $\X$ is a finite set of data points in high dimensional spaces or metric spaces. First, instead
of working with the level set of a single value which is difficult
to capture in discrete settings, Mapper considers the preimage of
subinterval $h^{-1}([a, b])$. Second, it replaces by clustering the contraction of connected components in continuous settings.  Specifically, the procedure of
Mapper used in this paper is as follows.

\begin{enumerate}
\item \textbf{Level-set formation}. Cover the range of $h:\X\to \R$ by a
set of subintervals which overlap in neighbors, {\it i.e.} $U_i=[a_i, b_i]$ with $U_i\cap U_{i+1} \neq \emptyset$ and $U_i\cap U_j \cap U_k=\emptyset$, and stratify $\X$ into level sets by taking inverse images $h^{-1}([a_i,b_i])$; 
\item \textbf{Clustering}. On each level set
$h^{-1}([a_i,b_i])$, construct the connected components or point clusters; 
\item \textbf{Graph representation}. Represent each component or cluster by a node. Add an edge between a node pair whenever they have nonempty intersection.
\end{enumerate}

Mapper thus returns an undirected graph representing the
connectivity information between data clusters across level sets
$h^{-1}([a_i,b_i])$. See the example in Figure~1 (b). Note that those degree-one nodes
lie in the intervals containing local minima/maxima and the branching
(degree-three) nodes lie in the intervals with saddle points, a sort of critical points. 

More generally, if the filter value range $\Omega$ takes some higher dimensional space or other topological spaces, Mapper may return a simplicial complex which is however not pursued in this paper. This construction can easily yield a multiresolution structure by choosing subintervals of different granularities, which helps handle noise. 

The key choice in Mapper will be the filter map $h:\X \to \Omega$. In fact, the name, \emph{Mapper}, was coined to emphasize the importance of choosing such a map. There is no universal scheme for this choice, which may vary from application to application. In \cite{SMC07} some examples are presented with the choice of density function and a certain eccentricity function measuring data depth as filters. In the following, we will discuss it in detail in the setting of biomolecular folding problems, with a particular example in RNA hairpin folding.

\subsection{Mapper Design in Biomolecular Folding}

Simulation data in biomolecular systems produces massive data in high dimensional space, and exhibits heterogeneity in distributions. The general procedure of Mapper above is adapted toward such challenges. The first crucial design is to construct filters based on conditional density functions estimated from data, which effectively enables us to focus on important local regions in configuration spaces and separate less populated pathways from the overwhelmed uninterested states. In clustering we choose the single-linkage method to capture possibly non-convex clusters. Below we give a detailed description on these particular implementations.

\smallskip

\subsubsection{Conditional Density Filters for Mapper} Our key construction of filters here is based on conditional density functions estimated from data, conditioning on the states of interests. For example, in the study of folding process we extract configurations from folding events and focus on the region close to folded states, while in unfolding process we draw samples from unfolding events and pay more attention to the zone around extended states. Simulation trajectories of those processes are often dominated by stochastic fluctuations around the initial states. It is near the target states that one may observe interesting structural information about pathways. The conditional density filters are chosen to reflect the localized free energy landscape around the states on pathways without being disturbed by off-pathway structures which are quite noisy and of high population in samples. 

Although the simulation data of biomolecular systems often lie in a high dimensional configuration space, the degree of freedom are much less due to the constraints and cooperation among atoms in folding process. It is often expected that the pathway samples are concentrated around some low dimensional manifolds which can be described by a relatively small number of intrinsic reaction coordinates \cite{KavPNAS06}. The existence of multiple pathways as in the example of this paper may lead to holes in such manifolds with nontrivial topology. Note that in the continuous case, the Reeb graph of a (unconditional) density function defined on the Euclidean space $\R^n$ turns out to be trivially a tree. However conditional density functions adopted here may restrict on interesting regions where the loops in the Reeb graph might shed light on the hole structures. Reconstructing the low-dimensional topology of densely sampled regions, thus may disclose the nature of multiple pathways. In theory, it is possible to efficiently recover the topology from samples of such low dimensional manifolds \cite{SmaleTop2}. In this paper, through conditional density filters we approach such manifolds via data level sets and extract some low-dimensional topological features which provide structural evidence on the existence of multiple pathways. 

Here we describe a general approach to construct conditional density filters, which will be specialized in the next section with the application to RNA hairpin folding.  
\begin{itemize}
\item Draw random samples $S\subseteq \X$ from the folding events. 
Choose importance weights on $S$, $w(x)\geq 0$, with higher values on interested states. 

\item Define the filter function by
\begin{equation} \label{eq:denfilter}
h(x) = - \log \frac{\sum_{y\in S}  w(y) K(x,y) }{\sum_{y\in S} w(y)} ,
\end{equation}
where the kernel function is defined by
\begin{equation} \label{eq:kernel}
K(x,y) = e^{-\frac{d^\beta (x,y)}{\alpha}},  
\end{equation}
where $d(x,y)$ is some distance function between configuration $x$ and $y$, $\alpha>0$ is the band-width, and $\beta>0$ is the exponent. For example the Euclidean distance with $\beta=2$ is used in case of Gaussian kernels, or the Hamming distance between structural contact maps with $\beta=1$ is used later in this paper. 

\item Resample from $S$ according to the new distribution, 
\[ p(x)= \frac{w(x)}{\sum_x w(x)}. \]
To avoid the normalization in large data sets, we can use the rejection method or extended \cite{Liu04}, {\it e.g.} a sequential Bernoulli experiments where a new configuration $x$ is accepted with probability $q(x)=\frac{w(x)}{\max\{w(x)\}}$.  
\end{itemize}
These configurations, together with the filter function \eqref{eq:denfilter}, will be the inputs of Mapper procedure shown in the last section.

Filter \eqref{eq:denfilter} assumes a density function in Boltzman form $f(x) = \frac{1}{Z} e^{-h(x)}$,  with partition function $Z = \sum_x e^{-h(x)}$. Thus up to a constant filter \eqref{eq:denfilter} approximates the free energy near the folded state. Since only order information of $h(x)$ will be used below, it leads to the same result choosing any monotone transform on $h$, {\it e.g.} $\sum_{y\in S}  w(y) K(x,y)$. Our construction is equivalent to a kernel density estimator which can be replaced by other methods \cite{Silver86}. 
 
\smallskip
 
\subsubsection{Level-set Formation in Mapper} To increase the robustness of Mapper allowing more errors in density estimation, we only use the order information of filter \eqref{eq:denfilter} to construct level sets.

\begin{itemize}
\item \textbf{Level-set formation}. Order the samples according to values of $h(x)$, and classify the samples into $m$ consecutive overlapping groups of equal or similar size, whose filter value ranges ${[a_i, b_i]}$ cover the range of $h$.  
\end{itemize}
Up to an arbitrary small perturbation, a real valued function $h:\X\to\R$ induces a linear order on samples. Therefore any monotone transform on $h(x)$, such as $c_1 \exp c_2 h(x)$, leads to the same level sets.

\smallskip

\subsubsection{Clustering in Mapper}
The graphical representation of Mapper depends on the choice of clustering methods. Mapper itself does not place any prerequisite on the clustering algorithm. In the study of biomolecular folding such as RNA hairpins, our purpose is to identify those connected components in free energy or density level sets, which might be of non-convex shapes and whose numbers are unknown to us beforehand. Single-linkage clustering is the simplest choice to meet those two features. 

\begin{itemize}
\item On each level set, construct a weighted graph, with nodes for configurations and edge weights as pairwise distances. 

\item Find a Minimal Spanning Tree (MST) of such a graph.

\item Find a threshold value for edges. We construct a histogram of MST edge weights with $k$ bins. Once some empty bins are found from top bins containing $p$ longest edges, we set the threshold to be the center of the first empty bin. Otherwise, set the threshold the maximal edge (diameter). 

\item Truncate the graph by breaking those edges greater than the threshold, dividing the graph into connected components. 

\item Prune those components of size no more than $q$. 
\end{itemize}

Single-linkage will separate those clusters where within each cluster two points can be joined by a path consisting of short edges, but relatively longer edges are required to merge the clusters. When we draw random samples from compact connected components in an Euclidean space, the distances between configurations within the same components will drop down to zero as the sample size grows. Hence the distances across components will be kept in the longest edges and can be separated from a large amount of short edges. Thresholding above tries to capture such a gap. Truncation may create several components/clusters of different sizes, where pruning helps reduce the noise and identify those dominant components. 

In the continuous setting, single-linkage clustering will consistently locate
those connected components when the samples are dense enough \cite{Hart81}. Such a feature makes it a desirable choice for Mapper \cite{SMC07}, as well as density cluster trees \cite{Stuetzle03,ZhouWong08}. However, in the latter part of this paper, we will meet a discrete configuration space, \emph{i.e.} the space of contact maps as undirected graphs. 
Thus we need to explain in what sense we extend the ``connected components'' in such a discrete setting.

Equipped with a metric, \emph{e.g.} Hamming distance, the discrete configuration set can be viewed as a weighted complete graph, where each node represents a structure and the weight of an edge is the distance between its endpoints. Single-linkage 
clustering firstly builds up a Minimal Spanning Tree (MST) of this graph 
and then truncate the MST by keeping the edges with the length less than a given threshold, which breaks the MST into several connected components or clusters. In this way, single-linkage compute the 
components where two nodes within a component are joined by a path consisting
of the short edges, but relatively longer edges are required to merge 
different components. 

One may also consider other clustering schemes, such as $k$-means, which is
widely used in clustering the configurations in the biomolecular folding simulations.
In contrast to single-linkage, $k$-means attempts to find the clusters such that
within cluster \emph{any} two nodes are connected by a short edge, rather than
by a path made up of short edges. Therefore, roughly speaking, $k$-means attempts
to find \emph{spherical-shape} clusters while single-linkage can discover
\emph{snake-shape} clusters. Both may provide useful but different kinds of information in
biomolecular folding problems. However $k$-means needs one to specify the number of 
clusters {\it a priori} while the single-linkage does not. This is a short-coming for $k$-means
since we don't have such information in advance.  So in this
paper, single-linkage is chosen as the basic scheme and $k$-means is only used in
comparative studies, when we already know the number of clusters from single-linkage. 
Other choice of clustering methods includes average-linkage,
complete-linkage, spectral clustering \cite{HasTibFri01,DingZha07}, {\it etc.}, which are however
not pursued in this paper.


\section{Results and Discussion} 
Recently, \cite{Greg08} performed Serial Replica Exchange
Molecular Dynamics (SREMD)\cite{HBP08,MEP00} simulations of the
GCAA tetraloop (5'-GGGCGCAAGCCU-3') on the Folding@home distributed computing platform.
The hairpin motif consists of a primarily Watson-Crick base-paired stem
capped with a loop of unpaired or non-Watson-Crick base-paired
nucleotides, as shown in Figure~2 (a).
Despite their simple structures, there is some debate over whether or not there are intermediate states in the folding of hairpins, {\it e.g.} see \cite{Ma07}. 

With the technique developed in this paper, we are able to disclose the structures of multiple intermediate states on the folding pathways, which in the first time provides structural evidence from computer simulations about RNA hairpin folding pathways. The biological implications of this discovery are discussed in detail by Bowman {\it et. al.} \cite{Greg08}. Here, we only focus on details of data analysis.

The RNA molecule examined here has $389$ atoms.
Including the solvent there are about $N=12,000$ atoms in the system, yielding $3N=36,000$ parameters. 
To reduce the dimensionality of this large space we chose to represent each configuration with a contact map. Contact maps can faithfully describe the base-pair interactions in the stem, which provides important structural information of RNA hairpin folding.
A contact map is a bit string specifying pairs of contacting residues that are not immediately adjacent in the sequence. Thus, even this coarse-grained space is $\R^{55}$. Following Bowman {\it et. al.} \cite{Greg08}, we define the \emph{native state}
as any conformation with all four stem base-pair contacts formed.
Each of these base-pair contacts is referred to as a native
contact.  For example, Figure 2 (a) shows a native state whose contact map model is illustrated in (b). An \emph{unfolding} event is defined as the set of
conformations between the first point with no contacts between any
two residues on opposite sides of the stem and the first preceding
point with four native contacts. A \emph{refolding} event is
defined as the set of conformations between the first point with
no contacts between any two residues on opposite sides of the stem
and the first subsequent point where the number of native contacts
is four.

\subsection{Structural Analysis by Mapper}
Mapper is an ideal
tool for such a problem due to the enormous size of the simulation dataset, the high
probability of non-convex states,  and the need to identify
folding intermediates with low populations relative to the folded
and unfolded states.
Application of Mapper to this data set revealed a number of intermediate states.

The data generated from SREMD simulations is normally dominated by
the folded and unfolded structures. For example, a typical
refolding trajectory starts from an unfolded state, undergoing a
significant period of stochastic fluctuation around that, then
proceeds gradually to the folded state. It is in
the neighborhood of folded states that interesting structural
information about folding pathways are exhibited. Therefore, in
the construction of the conditional density filters, we treat folding
and unfolding separately. In the study of folding pathways, we
take configurations from refolding events, and then weight heavily a neighborhood around the native states. However in the study of unfolding pathways, we sample from unfolding
events, and focus on a neighborhood of the unfolded states. 

The following parameters are used to produce the results in Figure~3. 
We use the Hamming distance $d_H(x,y)$ between a pair of contact maps in the conditional density function (Equation \eqref{eq:denfilter}) and choose $\alpha=\beta=1$ in kernel \eqref{eq:kernel}. For simplicity, the importance weights are set to one within the neighborhood of the state of interest and zero otherwise. In refolding events, we choose a neighborhood within 7-bit Hamming distance from the native state in Figure~2. In unfolding events, a neighborhood of the extended state is chosen as the set of configurations with no more than 6 non-adjacent contacts formed. In the level-set formation, the filter is divided into 8 levels of equal size with $25\%$ overlap. In the clustering, a histogram with 5 bins is used, with thresholding from top bins consisting largest $p=20\%$ edges and the cluster pruning size $q=2\%$ of the level sample size. More details on parameter tuning will be provided in the supporting information.

The graphical output of Mapper with such parameters shows distinct pictures about folding and unfolding pathways. 
Unfolding has a single dominant pathway characterized by unzipping from the end base-pair (Figure~3 (a)), while folding process has two dominant pathways, passing through either the formation of the closing base-pair or the end base-pair (Figure~3 (b)). 
Such an observation reveals a number of intermediate states in the folding process, which supports the multi-state hypothesis. It is interesting to notice in Figure~3 that conditional density filters seem good indicators of reaction coordinates, suggesting that the folding/unfolding processes start from the densest zone and become sparser as the reactions proceed.

\subsection{Kinetic Verifications}
Are the two pathways in refolding (Figure~3 (b)) are truly separate pathways or just the artifact of noise? This question can be answered from the kinetic information of simulation trajectories by computing the transition probability. Note that our purpose here is not to create a Markov model \cite{Chodera07} for metastable states, but investigate how the two intermediate states in refolding pathways are kinetically connected. Therefore the shortest lag time, 2~ps, is chosen which provides the finest resolution in simulation trajectories. 

To simply the result, we merge the four nodes with extended structures as a single unfolded state, U, and collapse the three blue nodes with folded structures as folded state, F, leaving alone the two intermediates, I1 and I2. This does not change the topology of Mapper graph, but highlights the dynamics associated with intermediate states. Configurations in simulations are mapped to such four-node states by nearest neighbor method. One-step (2 ps) transition probability are then computed among the four states. 

The result is shown in Figure~4. It can be seen that the two intermediates, I1 and I2, are kinetically well separated on folding pathways. Once the simulation climb up the energy barrier I1 and I2, the majority will either proceed to F or withdraw to U, while an ignorable minority will cross the intermediates from I1 to I2. 


\subsection{Importance of Conditional Density Filters}
Conditional density filters play a crucial role here, without which clustering methods like K-means or single-linkage tend to split the sparse intermediates and lump them with densest clusters. 

To see this, we make a comparison between Mapper clusters found in Figure~3 (b) and $K$-means clustering on the same data set. Since the number of $K$-means clusters is not unknown {\it a priori}, we performed a series of experiments with $k$ varying from 1 to 80, each of which has 20 repeated experiments. Our first purpose is to locate the value of $k$ around which the Mapper cluster with end base-pair formed becomes identifiable. Hence for each $K$-means experiment, we count the number of the end base-pair clusters, defined as the clusters containing more than $75\%$ configurations with native contact 4 (Figure~2 (b)) formed, and less than $25\%$ for any other native contact. Figure~5 plots a rough distribution of the numbers of end base-pair clusters against the growth of $k$. It can be seen that around $k=25$ this intermediate state becomes identifiable, in the sense that with more than 1/2 probability such clusters are found indicated by nonzero medians. Notice that as $k$ grows, the variation range ($10\%\sim 90\%$) of such cluster numbers expands, showing a trend of increasing instability. Particularly around $k=55$, such a state begins to split into several $K$-means clusters. 

We can further see how $K$-means clusters might split the intermediate states and lump them toward densest clusters. Figure~6 illustrates this when $k=30$ for $K$-means clustering, on the same data set for the construction of Mapper clusters on refolding pathways. 

\subsection{Comparative Studies on Single-linkage vs. $k$-means}
Single-linkage clustering is motivated by its ability to identify possibly non-convex clusters of unknown number. It is also interesting to explore other clustering methods such as K-means which tries to group data in \emph{spherical} clusters and is widely used in the studies of biomolecular folding simulations. Given the cluster number returned by single-linkage, comparisons with K-means of similar number of clusters on the same level sets might disclose how far the intermediate states deviate from spherical shapes. For this purpose, we perform K-means clustering on the same data set for refolding pathways in Figure 3 (b). We use the same number of clusters returned by single-linkage and especially on level five we set $k=2$. It turns out that K-means finds two clusters on level five with similar structural features to single-linkage, {\it i.e.} one with closing base-pair formed and the other with the end base-pair. However K-means has different partition: $48\%$ vs. $52\%$, in contrast to $23\%$ vs. $44\%$ in single-linkage. Clearly to form spherical clusters, K-means clusters mix more configurations from different single-linkage clusters, which can be shown by the percentage dropping of dominant end base-pair from $96\%$ to $65\%$ in the smaller cluster. However the structural similarity in both methods suggests that single-linkage clusters are not very far from spherical shapes. 

\subsection{On Nonlinear Dimensionality Reduction} 
Although a biomolecular system is typically described by a high dimensional configuration space, it is expected that those configurations often visited in a folding process may concentrate around some low-dimensional manifolds which might be described by a much smaller number of reaction coordinates. 
Recently Das {\it et. al.} \cite{KavPNAS06} shows that ISOMAP can be applied to recover such reaction coordinates in simple folding processes with a single pathway. Isomap tries to preserve both the local and global geodesic distance between configurations defined as shortest path distance on a neighborhood graph. However, ISOMAP might not work in complex problems where multiple pathways exist. ISOMAP requires that the data manifold are globally isometric to a convex domain of low dimensional space \cite{TDL00,Donoho03}. The existence of more than two pathways connecting two metastates, may lead to holes in sampled regions which fails the convex domain assumption. Moreover, ISOMAP is too sensitive to the metric in choice. In this paper we use a coarse metric as Hamming distance for contact maps, where the geodesic distance between configurations does not reflect the distance in folding process. Moreover, The high heterogeneity in distribution is also a hurdle for ISOMAP technique to identify useful intermediates.  

The last two issues also challenge other techniques for nonlinear dimensionality reduction, such as LLE \cite{RS00}, Laplacian Eigenmap \cite{BN03}, Hessian Eigenmap \cite{Donoho03}, and Diffusion map \cite{Coifman05}, {\it etc.} These geometric embedding techniques maps the data in high dimensional spaces to a low dimensional space by preserving some local metric relations among neighbors of data points, {\it e.g.} see \cite{JMS08}. They are thus sensitive to the metric in choice and heterogeneous distribution might distort local metrics. In applications to complex biomolecular systems, successful examples are only found in simple settings such as with a single protein folding pathway \cite{KavPNAS06} or quasi-steady state in dynamics of signal transduction networks \cite{GG07}. 
  
However, as a topological tool Mapper with density filters is shown efficient in dealing with heterogeneous distributions and less sensitive to the metric in choice. In this paper even with such a coarse metric as Hamming distance, it efficiently discloses structural information in pathways which are difficult to other geometric embedding techniques. Thus, one of our ongoing directions is to combine the topological tool Mapper with those geometric embedding techniques, such as applying nonlinear dimensionality reduction separately on components or clusters discovered by Mapper.

\section{Conclusions}
In this paper we develop Mapper, a topological data analysis tool, in the analysis of simulation data for biomolecular folding pathways. As an application, in the first time we are able to obtain structural evidence from computer simulations in support that RNA hairpin folding has two dominant pathways with multiple intermediate states. It is thus a promising direction to explore with Mapper such structural information in biomolecular folding problems. 

We have shown that with proper designs of conditional density filters and clustering schemes, Mapper can address the heterogeneity issue in distribution, deal with multiple pathway data with nontrivial topology, and be less sensitive to the metric in choice. These features can be used to enhance traditional nonlinear dimensionality reduction methods, such as ISOMAP, Laplacian Eigenmap, Diffusion maps, etc. One of our ongoing direction is to explore the combinations of the topological tool Mapper with those geometric tools for better characterizations of biomolecular systems.   

Only a limited use of data has been pursued for Mapper in this paper, where we ignore the kinetic information in simulation trajectories. Since we did not take into account the kinetic information in building up the model, the
intermediate states we identify are thermodynamically relevant states,
but not necessarily to be kinetically relevant states. Our future direction is to incorporate such kinetic information for developments of novel dynamical models. 

\section*{Acknowledgments}
YY would like to thank Wing-Hung Wong, Nancy Zhang, and Qing Zhou for helpful discussions, XH thanks Michael Levitt for his support. We also thank the world-wide Folding@home users for providing computing resources. 

\bibliographystyle{plain}

\newpage

\begin{figure}[h]
\centering
\centerline{\includegraphics[width=0.6\textwidth]{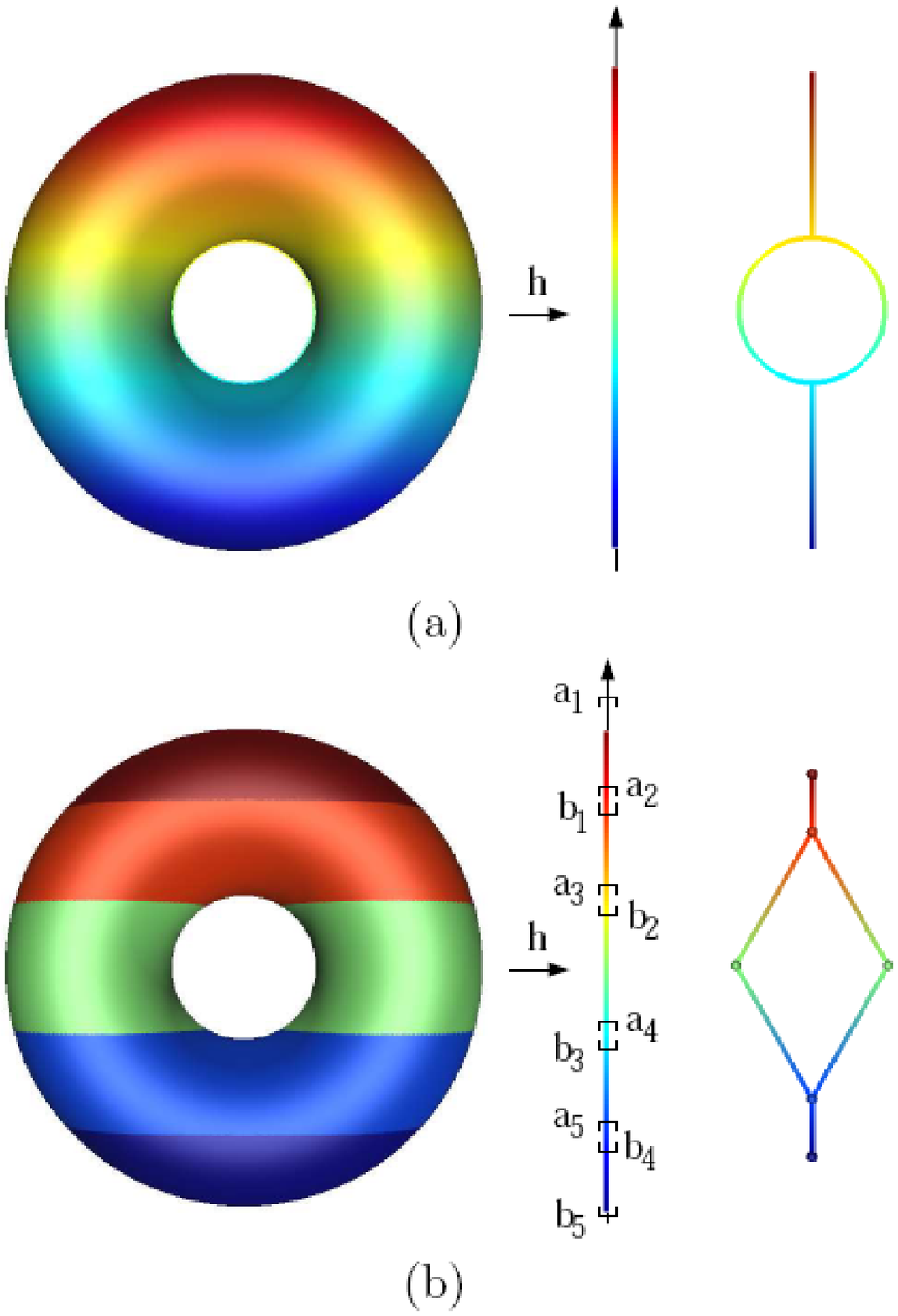}}
\caption{(a) Construction of Reeb graph; (b) Construction of Mapper. $h$ maps each point on torus to its height. \label{fig:torus}} 
\end{figure}


\begin{figure}[h] 
\begin{center}
\centerline{\includegraphics[width=0.6\textwidth]{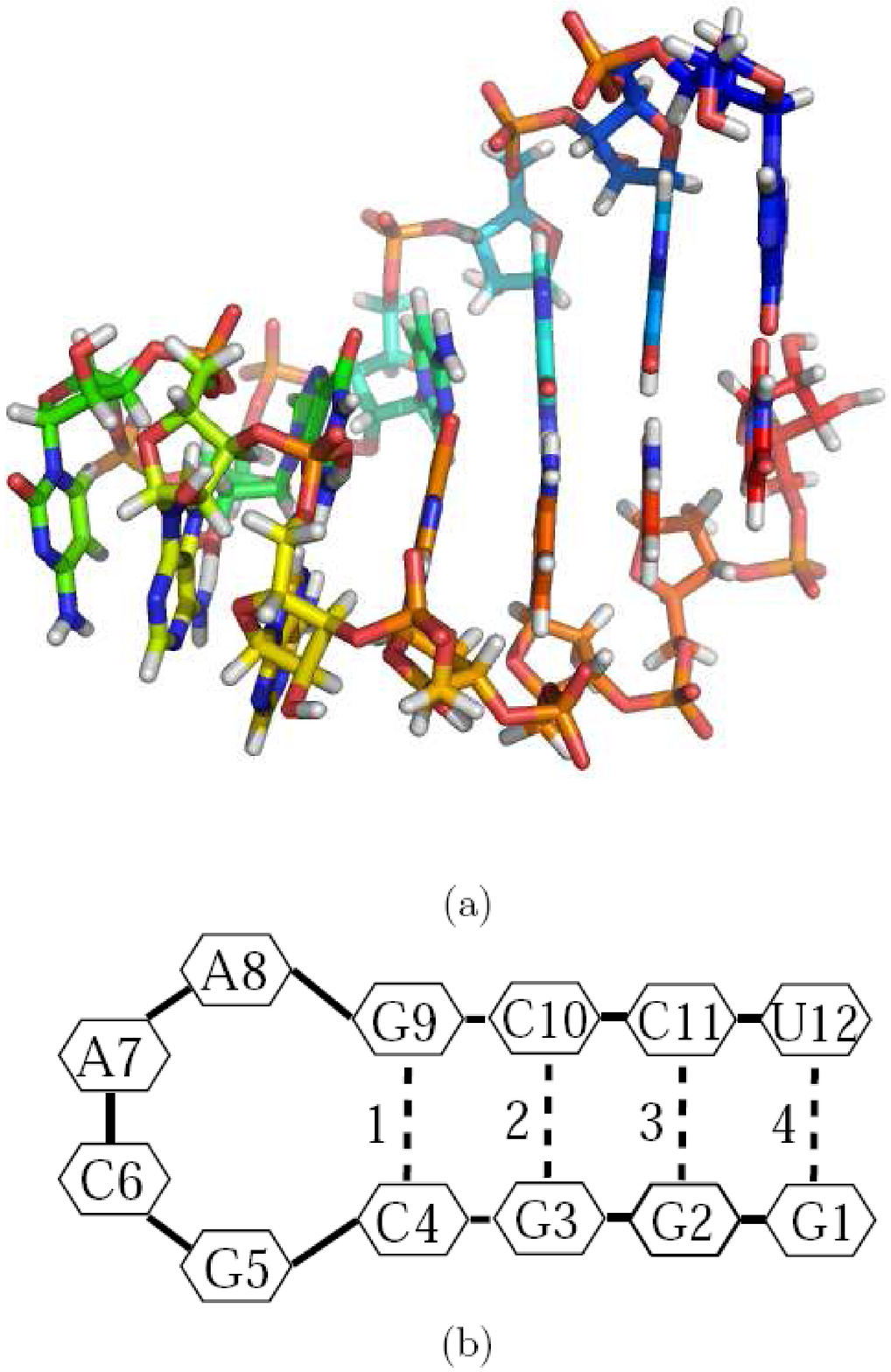}}
\caption{(a) NMR structure of the GCAA tetraloop.  (b) Contact map for the native state.
Bases are numbered from 1 to 12 and native basepair contacts (dotted lines) are numbered
1-4.\label{fig:native}}
\end{center}
\end{figure}


\begin{figure}[h] 
\begin{center}
\centerline{\includegraphics[width=0.6\textwidth]{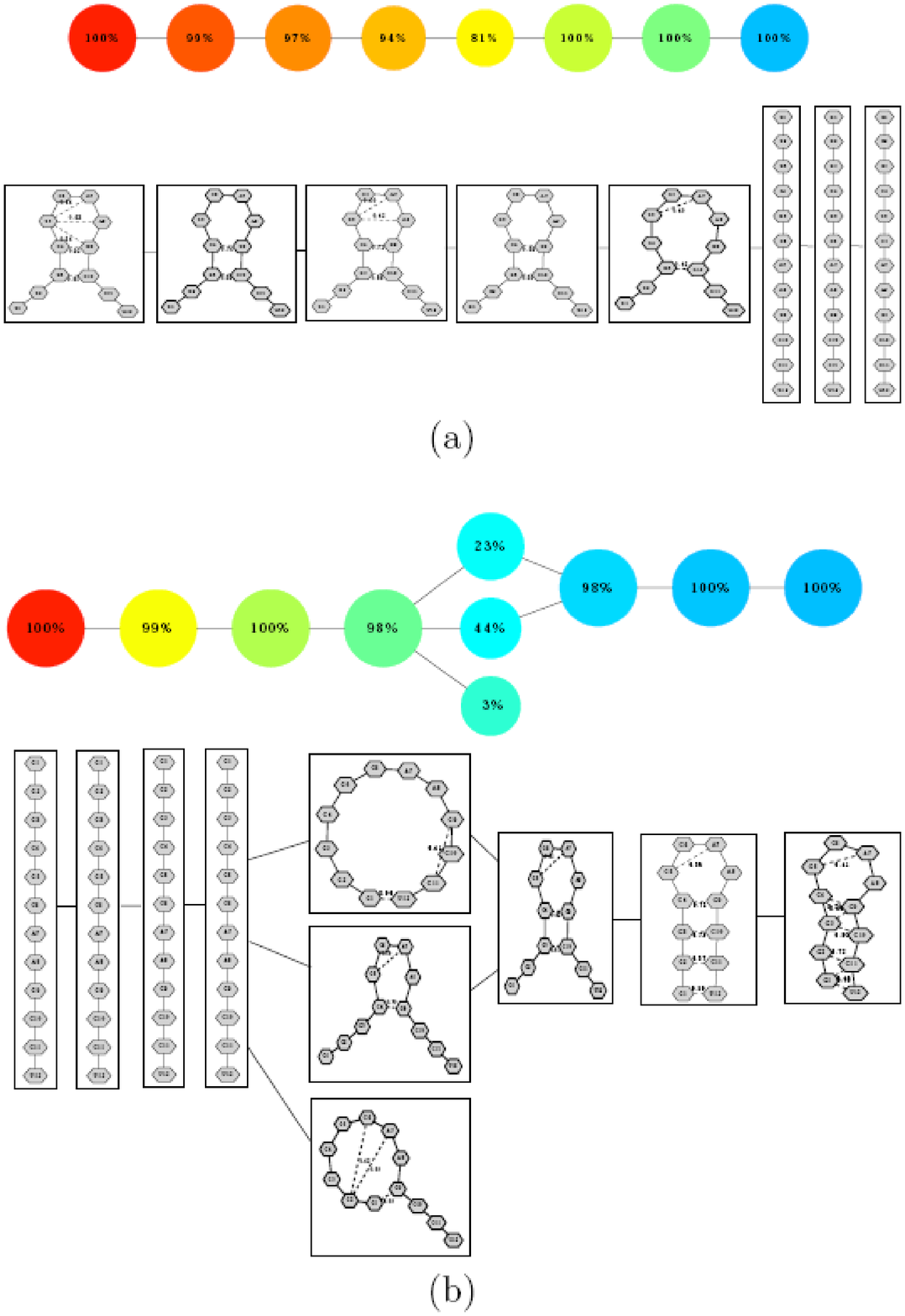}}
\caption{Graphical representation of pathways by Mapper. (a) Unfolding pathway. (b) Folding pathway. \label{fig:pathway} In both cases, the top row graphs are the outputs from Mapper, while the bottom row depicts the mean contact maps of the corresponding clusters. For clarity in mean contact maps we drop those mean contacts lower than $0.4$. The node colors from red to blue indicate the density from high to low, and the labels ({\it e.g.} $100\%$) show the percentage of configurations of the same level included in the cluster corresponding to the node. We dropped all the clusters of size smaller than $3\%$ of the level size. (a) shows that unfolding has a single dominant pathway characterized by unzipping from the end base-pair. (b) shows that folding process has two dominant pathways, passing through either the formation of the closing base-pair or the end base-pair. A noisy cluster consisting $3\%$ of the level size was also shown in (b), which accounts for reptation, {\it i.e.} sliding of the two strands of the stem. }
\end{center}
\end{figure}

\newpage 

\begin{figure}[h]
\begin{center}
\centerline{\includegraphics[width=\textwidth]{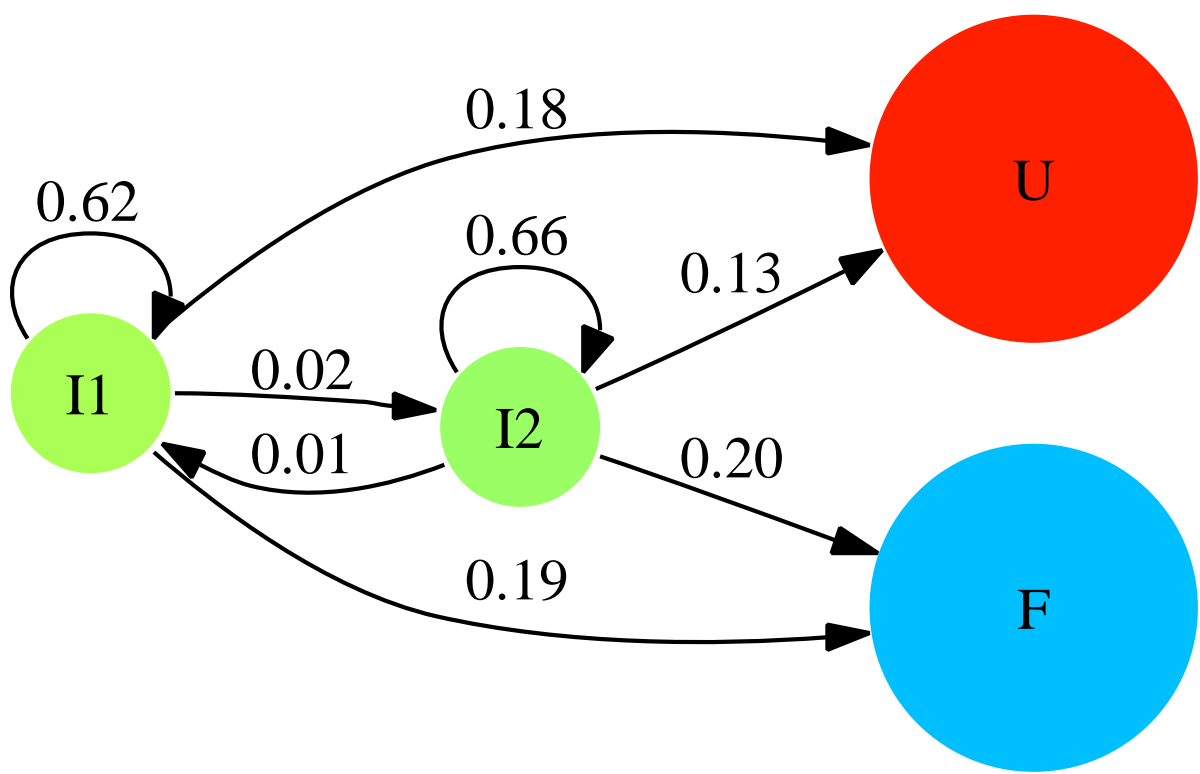}}
\caption{ \label{fig:transit4} Transition probability from two intermediate states. Lag time is $2ps$. The left four nodes as extended structures (Figure~3 (b)) are merged into node U, and the right three nodes as folded structures are collected in node F. The two intermediate states on pathways are denoted by I1 and I2, respectively. The transition probability from  I1 and I2 to other states are noted as numbers on the arrows. One can see that I1 and I2 are kinetically separated.}
\end{center}
\end{figure}


\begin{figure}[h]
\begin{center}
\centerline{\includegraphics[width=\textwidth]{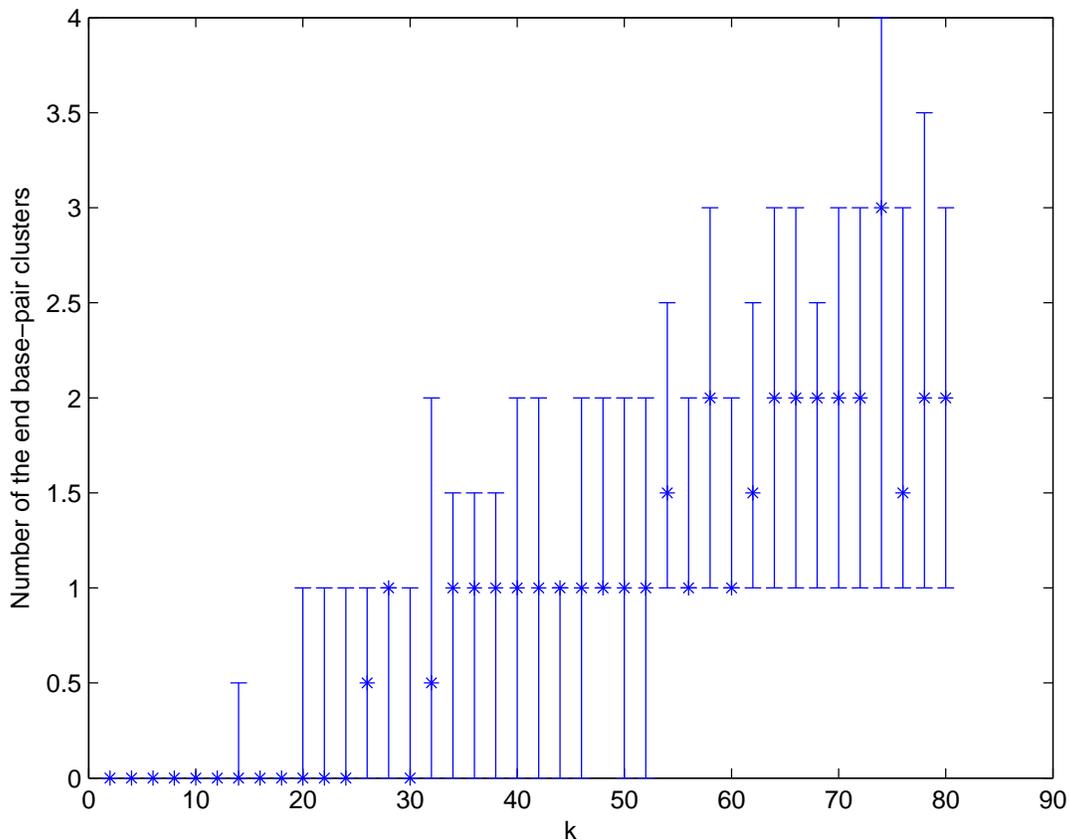}}
\caption{ \label{fig:k80} The number of end base-pair clusters found by $K$-means. Here $k$ ranges from 2 to 80 with step 2. For each $k$, 20 experiments are repeated with $K$-means clustering. The number of clusters with end base-pair formed are recorded. The star is the median of such numbers and the bar delimits the distribution range from $10\%$ to $90\%$. Starting from around $k=25$, such clusters appear with at least 1/2 probability. Around $k=55$, such clusters begin to split. The instability of $K$-means clusters is increasing as $k$ grows, indicated by the expanding ranges.}
\end{center}
\end{figure}


\begin{figure}[h]
\begin{center}
\centerline{\includegraphics[width=0.6\textwidth]{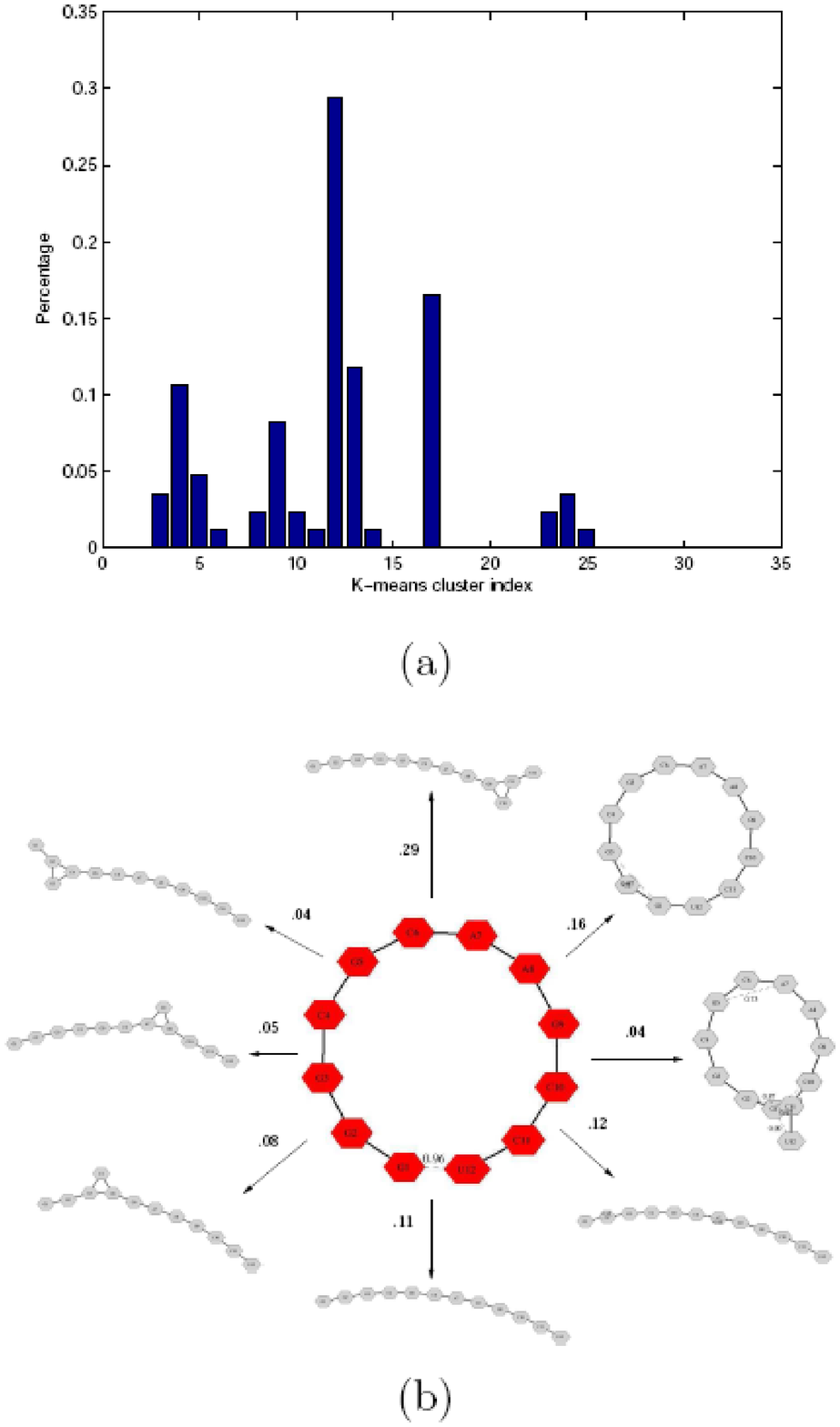}}
\caption{ \label{fig:confusion} K-means clustering fails to capture the low density intermediate
states with one end base-pair formed.  The illustration here chooses $k=30$ for $K$-means clustering. (a) shows how end base-pair formed structures are distributed in different k-means clusters; (b) illustrates the mean structures of the top eight K-means clusters (gray) which contain base-pair formed structures.
$K$-means splits the Mapper cluster and lumps them with densest clusters.}
\end{center}
\end{figure}

\end{document}